\RequirePackage{ifpdf}
\ifpdf 
\documentclass[pdftex]{sigma}
\else
\documentclass{sigma}
\fi

\numberwithin{equation}{section}

\newcommand{\sn}{\mathop{\rm sn}\nolimits}
\def\cal{\fam2}

\begin{document}

\renewcommand{\PaperNumber}{053}

\FirstPageHeading

\renewcommand{\thefootnote}{$\star$}

\ShortArticleName{Extended Reduced Ostrovsky Equation}

\ArticleName{Periodic and Solitary Travelling-Wave Solutions\\ of an
Extended Reduced Ostrovsky Equation\footnote{This
paper is a contribution to the Proceedings of the Seventh
International Conference ``Symmetry in Nonlinear Mathematical
Physics'' (June 24--30, 2007, Kyiv, Ukraine). The full collection
is available at
\href{http://www.emis.de/journals/SIGMA/symmetry2007.html}{http://www.emis.de/journals/SIGMA/symmetry2007.html}}}

\Author{E. John PARKES}

\AuthorNameForHeading{E.J. Parkes}

\Address{Department of Mathematics, University of
Strathclyde, Glasgow G1 1XH, UK}
\Email{\href{mailto:ejp@maths.strath.ac.uk}{ejp@maths.strath.ac.uk}}
\URLaddress{\url{http://www.maths.strath.ac.uk/~caas35/}}

\ArticleDates{Received October 29, 2007, in f\/inal form June 16,
2008; Published online June 19, 2008}

\Abstract{Periodic and solitary travelling-wave solutions of an extended reduced Ostrovsky equation
are investigated.
Attention is restricted to solutions that, for the appropriate choice
of certain constant parameters,
reduce to solutions of the reduced Ostrovsky equation.
It is shown how the nature of the waves may be categorized in a simple
way by considering the value of a certain single combination of constant parameters.
The periodic waves may be smooth humps, cuspons, loops
or parabolic corner waves.
The latter are shown to be the maximum-amplitude limit
of a one-parameter family
of periodic smooth-hump waves.
The solitary waves may be a smooth hump, a cuspon, a loop or a parabolic
wave with compact support. All the solutions are expressed in parametric form.
Only in one circumstance can the variable parameter be eliminated to give a solution in explicit form.
In this case the resulting waves are either a solitary parabolic wave with compact support or
the corresponding periodic corner waves.}


\Classification{35Q58; 35Q53: 35C05}

\section{Introduction}\label{sec1}

The equation
\begin{gather}
(u_t+c_0u_x+ a uu_x+ b u_{xxx})_x=\gamma u,
\label{OE}
\end{gather}
where $c_0$ is the velocity of dispersionless linear waves, $a$ is the
nonlinear coef\/f\/icient, and $b$ and~$\gamma$ are dispersion coef\/f\/icients,
is a model for weakly nonlinear surface and internal waves in a~rotating ocean. It was derived by Ostrovsky in 1978 \cite{O78} and is
now known as the Ostrovsky equation.
For long waves, for which high-frequency dispersion is negligible, $b=0$
and (\ref{OE}) becomes the so called reduced Ostrovsky equation (ROE),
namely
\begin{gather}
(u_t+c_0u_x+a uu_x)_x=\gamma u.
\label{ROE}
\end{gather}
The ROE has been studied by several authors (see \cite{S06}
and references therein).

By applying the transformation
\begin{gather}
u\to u/a, \qquad t\to t/\sqrt{|\gamma|}, \qquad x\to (x+c_0t)/\sqrt{|\gamma|}
\end{gather}
to (\ref{ROE}), we obtain the ROE in the neat form
\begin{gather}
\frac{\partial}{\partial x}\,{\cal{D}}u+\delta u=0,
\qquad \hbox{where}\qquad {\cal{D}}:=\frac{\partial}{\partial t}+u\frac{\partial}{\partial x}
\qquad\hbox{and}\qquad \delta:=-\frac{\gamma}{|\gamma|}=\pm 1.
\label{simpleROE}
\end{gather}
In \cite{P07} we
found periodic and solitary travelling-wave solutions of (\ref{simpleROE}).
(By a travelling-wave solution we mean one in which the dependence on $x$ and $t$
is via a single variable
$\chi:=x-vt-x_0$, where $v$ and $x_0$ are arbitrary constants.)

As mentioned in \cite{B05EJAM} and
references cited therein, equation (\ref{simpleROE}), with $\delta=-1$, is sometimes referred
to as the Ostrovsky--Hunter equation (OHE).
Vakhnenko derived equation (\ref{simpleROE}), with $\delta=1$, in order
to model the propagation of waves in a relaxing medium \cite{Va92,V99}.
Parkes \cite{Pa93} dubbed this equation the Vakhnenko equation (VE).

In \cite{P07} we pointed out that equation (\ref{simpleROE}) is invariant under the transformation
\begin{gather}
u\to -u, \qquad t\to -t, \qquad \delta\to -\delta
\end{gather}
so that the solutions of the OHE and VE are related in a simple way.
For example, given a travelling-wave solution to one of the equations,
the corresponding solution to the other equation is the inverted wave travelling in the
opposite direction.

When $v>0$ the travelling-wave solutions of the OHE are periodic smooth-hump waves
\cite[Fig.~1]{P07} which, in the limit of maximum amplitude, become `corner waves'
\cite[Fig.~2]{P07}; the latter have discontinuous slope at each crest and
are parabolic between crests. The corresponding solutions for the VE occur when $v<0$,
namely periodic smooth-hump waves and parabolic corner waves with discontinuous slope
at the troughs \cite[Fig.~2]{Va92}. When $v<0$ the travelling-wave solutions of the OHE
are periodic inverted loops \cite[Fig.~3]{P07} and a solitary inverted loop \cite[Fig.~4]{P07}.
The corresponding solutions for the VE occur when $v>0$, namely periodic loops and a solitary loop
\cite[Fig.~1]{Va92}.

The corner wave solution for the OHE
was discussed in detail by Boyd \cite{B05EJAM} and Parkes \cite{P07}.
Clearly a similar discussion applies to the corner-wave solution of the VE.

In \cite{MoPaVa99} we showed that the solitary-loop solution of the VE
when $v>0$ is a soliton,
and that the VE has a multi-soliton solution in which each soliton is a
loop that propagates in the positive
$x$-direction. During interaction the loop-solitons combine in a rather remarkable way as is
illustrated in Figs.~3--5
in \cite{MoPaVa99} for example. Clearly, similar observations apply to the inverted
solitary-loop solution of the OHE when $v<0$.

In \cite{MoPa03} we considered a Hirota--Satsuma-type `shallow water wave'
equation \cite{EsFu94} of the form
\begin{gather}
U_{XXT}+pUU_T-qU_X\int^{\infty}_X U_T(X',T)\,dX'+\beta U_T+qU_X=0,
\label{Uequation}
\end{gather}
where $p\ne 0$, $q\ne 0$ and $\beta$ are arbitrary constants. By using the
transformation
\begin{gather}
x=T+\int^X_{-\infty} U(X',T)\,dX'+x_0,\qquad t=X, \qquad u(x,t)=U(X,T),
\label{XTtrans}
\end{gather}
where $x_0$ is a constant, we obtained the following equation:
\begin{gather}
\frac{\partial}{\partial x}\left({\cal D}^2u+\frac{1}{2}\,pu^2+\beta u\right)+q{\cal D}u=0.
\label{exROE}
\end{gather}
With $p=q$ and $\beta=0$, (\ref{exROE}) may be written
\begin{gather}
\left(\frac{\partial u}{\partial x}
+{\cal D}\right)\left( \frac{\partial}{\partial x}\,{\cal D}u + pu\right)=0.
\label{diffROE}
\end{gather}
Clearly solutions of the ROE are also solutions to equation~(\ref{diffROE}) with $p=\pm 1$.
Because of this, hereafter we shall refer to the more general form of equation~(\ref{diffROE}),
namely equation~(\ref{exROE}), as the extended ROE (exROE).

In \cite{MoPa03} we explained why the exROE (with $\beta\ne 0$) is integrable in
two special cases, namely when $p=q$ and when $p=2q$.
In \cite{MoPa01} we considered the exROE with $p=q=1$ and $\beta\ne 0$.
We referred to the resulting equation as the generalised VE (GVE) and went on
to f\/ind its $N$-soliton solution. The $i$th soliton may be a hump, loop or cusp depending on
the value of $\beta/k_i^2$,
where $6k_i^2$ is the amplitude of the $i$th soliton.
In \cite{MoPa03} we considered the exROE with $p=2q$ and $\beta\ne 0$.
We referred to the resulting equation as the modif\/ied generalised VE (mGVE) and went on
to f\/ind its $N$-soliton solution. In \cite{MoPa03} we assumed that $q>0$
and then the $i$th soliton may be a~hump, loop or cusp depending on
the value of $\beta/k_i^2$,
where $4k_i^2/q$ is the amplitude of the $i$th soliton. For $q<0$, the
three possible types of soliton are the inverted versions of those for $q>0$.

In \cite{MoPa03,MoPa01} we considered two-soliton interactions, i.e~the case $N=2$.
We found that for both the GVE and the mGVE, `hump-hump', `loop-loop' and `hump-loop'
2-soliton interactions are possible. In addition,
`cusp-loop' and `hump-cusp' interactions are possible for the GVE,
and a~`cusp-cusp' interaction is possible for the mGVE.

Liu et al \cite{LLW07} investigated periodic and solitary travelling-wave solutions of the exROE.
Their f\/irst step was to follow the procedure described in \cite{MoPa03}, namely to introduce
new independent variables $X$ and $T$ as def\/ined in (\ref{XTtrans})
and hence to transform the exROE into equation (\ref{Uequation}).
Then they used the Jacobi elliptic-function expansion
method (see \cite{PaDuAb02}, for example) to
f\/ind a solution to equation (\ref{Uequation}) in terms of the elliptic sn function.
A transformation back to the original independent variables leads to implicit
periodic and solitary-wave solutions of the exROE.

The f\/irst aim of the present paper is to f\/ind implicit periodic and solitary travelling-wave
solutions of the exROE that have the property that they reduce to the
bounded solutions of the ROE for the appropriate
choice of parameters, namely $p=q=\pm 1$ and $\beta=0$. This consideration,
together with the fact that bounded solutions of the ROE have $v\ne 0$,
lead us to seek solutions of the exROE subject to the restrictions
\begin{gather}
p+q\ne 0, \qquad qv-\beta\ne 0, \qquad B=0,
\label{restrictions}
\end{gather}
where $B$ is a constant of integration that is def\/ined in Section~\ref{sec2}.

The solution procedure that we adopt is the one that we have used previously to f\/ind
implicit periodic and solitary travelling-wave solutions of the ROE \cite{P07},
the Degasperis--Procesi equation~\cite{VP04},
the Camassa--Holm equation~\cite{PV05} and the short-pulse equation~\cite{P08}.
An important feature of the method is that it delivers solutions in which both the dependent variable
and the independent variable~$\chi$
are given in terms of a parameter. It may or may not be possible to eliminate the parameter
in order to obtain an explicit solution in which the dependent variable is given
explicitly in terms of $\chi$.
Qiao et al. have tackled similar problems but have restricted
attention only to explicit solutions for solitary waves for which the dependent
variable tends to a~constant as~$|x|\to\infty$;
see, for example, \cite{ZQ07,Q08} for the Degasperis--Procesi equation
and \cite{QZ06} for the Camassa--Holm equation. (Related aspects of the Camassa--Holm hierarchy are
discussed in~\cite{Q03}.) We claim that Qiao's method does not yield such a solution for the problem
considered in this paper; the only explicit solitary-wave solution
derived by our method is a wave with compact support.

Our solution procedure is quite dif\/ferent from the one used in~\cite{LLW07}.
After correcting some minor errors in \cite{LLW07}, we will show that the solutions in~\cite{LLW07}
agree with our results. Liu et al~\cite{LLW07} mention that
their solutions may be of dif\/ferent types such as loops, humps or cusps,
but they made no attempt to categorize the solutions according to appropriate parameter ranges.
The second aim in the present paper is to provide such a categorization.

In Section~\ref{sec2} we f\/ind that the quest for travelling-wave solutions of the exROE leads to
a simple integrated form of equation (\ref{exROE}).
As this is similar in form to the corresponding
equation for the VE, in Section~\ref{sec3} we use known results for the VE to generate implicit
periodic and solitary-wave solutions to the exROE. In Section~\ref{sec4} we categorize these solutions
according to the shape of the corresponding wave prof\/ile. In Section~\ref{sec5} we illustrate our results
with two examples.
Some conclusions are given in Section~\ref{sec6}.
In Appendix A we point out some errors in~\cite{LLW07}.
In Appendix B we indicate how single-valued composite solutions may be obtained from the results for
multi-valued solutions derived in this paper.

\section{An integrated form of the exROE}\label{sec2}

As explained in Section~\ref{sec1}, here, and subsequently, we assume that $p+q\ne 0$ and $qv-\beta\ne 0$.
In order to seek travelling-wave solutions of equation~(\ref{exROE}),
it is convenient to introduce a new dependent
variable $z$ def\/ined by
\begin{gather}
z=\frac{(p+q)u}{2|qv-\beta|}-c,\qquad\hbox{where}\qquad
c:=\frac{qv-\beta}{|qv-\beta|}=\pm 1,
\label{z}
\end{gather}
and to assume that $z$ is an implicit or explicit function of $\eta$, where
\begin{gather}
\eta=\frac{\chi}{|qv-\beta|^{1/2}}.
\label{eta}
\end{gather}
It is also convenient to introduce the variable $\zeta$ def\/ined by the relation
\begin{gather}
\frac{d\eta}{d\zeta}=\frac{u-v}{|qv-\beta|}.
\label{zeta}
\end{gather}
(Note that $\zeta$ is not a new spatial variable; it is the parameter in the parametric
form of solution that we obtain eventually.)
Then (\ref{exROE}) becomes
\begin{gather}
z_{\zeta\zeta\zeta}+2zz_{\zeta}+cz_{\zeta}=0.
\label{exROEode}
\end{gather}
After one integration, (\ref{exROEode}) gives
\begin{gather}
z_{\zeta\zeta}+z^2+cz=B,
\label{VEodeB}
\end{gather}
where $B$ is a constant of integration.

We impose the requirement that, for $p=q=1$ and $\beta=0$,
the solutions that we seek reduce
to the corresponding solutions of the VE.
To do this we note that, with $p=q=1$ and $\beta=0$,
(\ref{z}) and (\ref{eta}) give
\begin{gather}
z=\frac{u-v}{|v|} ,\qquad \eta=\frac{\chi}{|v|^{1/2}} ,
\label{VEzandeta}
\end{gather}
and then (\ref{zeta}) gives
\begin{gather}
\frac{d\eta}{d\zeta}=z.
\label{zetaVE}
\end{gather}
Now (\ref{VEodeB}) and (\ref{zetaVE}) give
\begin{gather}
z(zz_{\eta})_{\eta}+z^2+cz=B,\qquad\hbox{where}\qquad c:=\frac{v}{|v|}=\pm 1.
\label{VEode}
\end{gather}
The corresponding relations for the VE are given in \cite{P07}; (\ref{VEzandeta})
agrees with (2.6) in \cite{P07}, and (\ref{VEode}) with $B=0$ agrees with (2.7) in \cite{P07}.
Accordingly we set $B=0$ from here on.

With $B=0$, equation~(\ref{VEodeB}) can be integrated once more to give
\begin{gather}
z_{\zeta}^2=f(z):=-\frac{2}{3}z^3-cz^2+\frac{1}{3}c^3A,
\label{poteqn}
\end{gather}
where $A$ is a real constant.
Equation (\ref{poteqn}) is one of the dif\/ferential equations that arise in solving the VE;
it is equivalent to equations~(2.9) and (2.10) in \cite{P07} with $\delta=1$.
We make use of this in Section~\ref{sec3}.
Also, as noted in \cite{P07},
the cubic equation $f(z)=0$ has three real roots provided that $0\le A\le 1$.

\section{Travelling-wave solutions of the exROE}\label{sec3}

The bounded solutions of equation~(\ref{poteqn}) that we seek are such that $z_1\le z_2 \le z \le z_3$,
where $z_1$, $z_2$ and $z_3$ are the three real roots of $f(z)=0$. In \cite[Appendix]{P07}
we gave expressions for these roots and $m:=(z_3-z_2)/(z_3-z_1)$ in terms of an angle $\theta$.
By eliminating $\theta$, we obtain $z_1$, $z_2$ and $z_3$ in terms of $m$, namely
\begin{eqnarray}
z_1&=&-\frac{c}{2}+\frac{m-2}{2\sqrt{m^2-m+1}} ,\label{z1}\\
z_2&=&-\frac{c}{2}+\frac{1-2m}{2\sqrt{m^2-m+1}} ,\label{z2}\\
z_3&=&-\frac{c}{2}+\frac{1+m}{2\sqrt{m^2-m+1}} ,\label{z3}
\end{eqnarray}
where $0\le m\le 1$.

Following \cite[Section 3]{P07}, we may integrate equation~(\ref{poteqn}) by using result 236.00 in
\cite{BF71} to obtain
\begin{gather}
z=z_3-(z_3-z_2)\sn^2(w|m),\qquad\hbox{where}\qquad w=\sqrt{\frac{z_3-z_1}{6}}\,\zeta.
\label{VEgensol}
\end{gather}
Result 310.02 in \cite{BF71} leads to
\begin{gather}
\int\! z\,dw=z_1w+(z_3-z_1)E(w|m)+\hbox{const}.
\label{VEeta}
\end{gather}
In (\ref{VEgensol}) $\sn(w|m)$ is a Jacobian elliptic function and the notation is
as used in \cite[Chapter 16]{AS72}; in (\ref{VEeta}) $E(w|m)$ is the elliptic integral
of the second kind and the notation is as used in \cite[Section 17.2.8]{AS72}.

In view of the def\/inition of $c$ in (\ref{z}), it is convenient to let
\begin{gather}
qv-\beta=4c\kappa^2,
\label{kappa}
\end{gather}
where $\kappa$ is a positive constant.
It is also convenient to def\/ine the positive constant $k$ by
\begin{gather}
\kappa^2=k^2\sqrt{m^2-m+1} .
\label{k}
\end{gather}

By using (\ref{z1})--(\ref{k}) in (\ref{z})--(\ref{zeta}), we obtain
\begin{gather}
u=\frac{4k^2}{p+q}\left[m+1+c\sqrt{m^2-m+1}-3m\sn^2(w|m)\right],
\label{u}
\end{gather}
\begin{gather}
\chi=\frac{4k}{p+q}\left[(m-2+c\sqrt{m^2-m+1})w+3E(w|m)\right]-\frac{(\beta+4c\kappa^2)w}{kq} .
\label{x}
\end{gather}
The travelling-wave solution to the exROE is given in parametric form by
(\ref{u}) and (\ref{x}) with $w$ as the parameter, so that $u$ is an implicit function of $\chi$.
This solution agrees with the corrected versions of (3.25) and (3.27) in \cite{LLW07}.
(We discuss the corrections in Appendix A.)
With respect to $w$, $u$ in (\ref{u}) is periodic with period $2K(m)$, where
$K(m)$ is the complete elliptic integral of the f\/irst kind. It follows from (\ref{x})
that the wavelength $\lambda$ of $u$ regarded as an implicit function of $\chi$ is
\begin{gather}
\lambda=
\left|\frac{8k}{p+q}\left[(m-2+c\sqrt{m^2-m+1})K(m)+3E(m)\right]-
\frac{2(\beta+4c\kappa^2)K(m)}{kq} \right|,
\label{wavelength}
\end{gather}
where $E(m)$ is the complete elliptic integral of the second kind.

When $m=1$, the solution given by (\ref{u}) and (\ref{x}) becomes
\begin{gather}
u=\frac{4k^2}{p+q}\left[2+c-3\tanh^2 w\right],
\label{usol}
\end{gather}
\begin{gather}
\chi=\frac{4k}{p+q}\left[(c-1)w+3\tanh w\right]-\frac{(\beta+4ck^2)w}{kq} .
\label{xsol}
\end{gather}
This solution agrees with the corrected versions of (3.26) and (3.28) in \cite{LLW07}.

\section{Categorization of solutions}\label{sec4}

In this section we categorize solutions according to
the shape of the corresponding wave prof\/ile.
We discuss the cases for which $0<m<1$ and $m=1$ separately.
Firstly we present some preliminary results.

\subsection{Preliminaries}\label{sec4.1}

In this sub-section we assume that $0<m<1$.

It is convenient to def\/ine a quantity $\psi$ by
\begin{gather}
\psi(m):=\frac{(p+q)\beta}{4q\kappa^2}+\frac{pc}{q} .
\label{psi}
\end{gather}
Note that by eliminating $\beta$ between (\ref{psi}) and (\ref{kappa}),
we obtain
\begin{gather}
(p+q)v=4\kappa^2(\psi+c),
\label{v}
\end{gather}
and by eliminating $\kappa^2$ between (\ref{psi}) and (\ref{kappa}),
we obtain
\begin{gather}
\psi=\frac{pv+\beta}{|qv-\beta|} .
\label{altpsi}
\end{gather}
Observe that, for a given choice of $p$, $q$, $\beta$ and $v$,
$\psi$ and $\kappa$ are constants independent of $m$, but $k$ depends on $m$;
on the other hand, for a given choice of $p$, $q$, $\beta$ and $k$, $\psi$,
$\kappa$ and $v$ depend on $m$.

From (\ref{x}) we have
\begin{gather}
\frac{d\chi}{dw}=\frac{4k}{p+q}\left[m+1+c\sqrt{m^2-m+1}-3m\sn^2(w|m)\right]
-\frac{(\beta+4c\kappa^2)}{kq}
\label{chideriv}
\end{gather}
which is a periodic function of $w$.
From (\ref{chideriv}) and (\ref{k})
we f\/ind that the maximum value of $(p+q)\frac{d\chi}{dw}$ is zero when
\begin{gather}
\psi=\psi_1(m):=\frac{(m+1)}{\sqrt{m^2-m+1}} .
\label{psi1}
\end{gather}
For $\psi>\psi_1(m)$, $(p+q)\chi$ is a strictly monotonic decreasing function of $w$.
Similarly the minimum value of $(p+q)\frac{d\chi}{dw}$ is zero when
\begin{gather}
\psi=\psi_3(m):=\frac{(1-2m)}{\sqrt{m^2-m+1}} .
\label{psi3}
\end{gather}
For $\psi<\psi_3(m)$, $(p+q)\chi$ is a strictly monotonic increasing function of $w$.

For $\psi_3(m)<\psi<\psi_1(m)$, $(p+q)\frac{d\chi}{dw}$ changes sign periodically.
At the values of $\chi$ where $\frac{d\chi}{dw}=0$, i.e.~where $u=v$
as can be seen from (\ref{zeta}), the prof\/ile of $u$ (regarded as an implicit function of $\chi$)
has inf\/inite slope.

Finally, we note that $\chi$ is periodic if $\lambda$ in (\ref{wavelength}) is zero;
this condition gives
\begin{gather}
\psi=\psi_2(m):=\frac{1}{\sqrt{m^2-m+1}}\left[m-2+\frac{3E(m)}{K(m)}\right].
\label{psi2}
\end{gather}
In this case, $\chi$ may be written
\begin{gather}
\chi=\frac{12k}{p+q}\left[E(w|m)-\frac{E(m)}{K(m)}w\right].
\label{periodicchi}
\end{gather}
The range of $\chi$ is $[-\chi_m,\chi_m]$, where
\begin{gather}
\chi_m=\frac{12k}{|p+q|}\left[E(w_m|m)-\frac{E(m)}{K(m)}w_m\right]
\label{chimrange}
\end{gather}
and $w_m$ is such that $0<w_m<K(m)$ and is given by
\begin{gather}
w_m=\sn^{-1}\sqrt{\frac{1}{m}\left[1-\frac{E(m)}{K(m)}\right]} .
\label{wm}
\end{gather}

\begin{figure}[t]
\centerline{\includegraphics[scale=0.85]{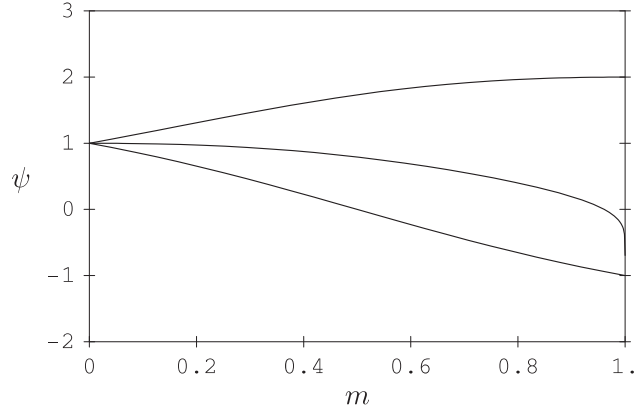}}
\vspace{-3mm}

\caption{The upper, middle and lower curves are $\psi=\psi_1(m)$,
$\psi=\psi_2(m)$ and $\psi=\psi_3(m)$, respectively.}\label{fig1}
\vspace{-3mm}
\end{figure}

The curves $\psi=\psi_1(m)$,
$\psi=\psi_2(m)$ and $\psi=\psi_3(m)$ are plotted in Fig.~\ref{fig1}.

\subsection[Waves with $0<m<1$]{Waves with $\boldsymbol{0<m<1}$}\label{sec4.2}

For $0<m<1$, the travelling-wave solution of the exROE is given by (\ref{u}) and (\ref{x}).
From~(\ref{u}) it can be seen that $u$ as a function of $w$ has a periodic smooth hump prof\/ile.
The nature of the corresponding prof\/ile of $u$ as a~function of $\chi$ clearly depends
on the behaviour of $\chi$ as a function of $w$ as given by (\ref{x}).
For all values of $\psi$ with $0<m<1$, except for $\psi=\psi_2(m)$,
the range of~$\chi$ as a function of $w$ is $(-\infty,\infty)$;
the corresponding possible periodic-wave prof\/iles for $(p+q)u$ may be categorized as follows:
\begin{center}
\begin{tabular}{ll}
$\psi>\psi_1(m)$:& smooth humps\\
$\psi=\psi_1(m)$:& cuspons\\
$\psi_2(m)<\psi<\psi_1(m)$:& loops\\
$\psi_3(m)<\psi<\psi_2(m)$:& inverted loops\\
$\psi=\psi_3(m)$:& inverted cuspons\\
$\psi<\psi_3(m)$:& smooth humps\\
\end{tabular}
\end{center}
Furthermore, from (\ref{v}), $(p+q)v\gtrless 0$ according as $\psi\gtrless -c$.

When $\psi=\psi_2(m)$, $\chi$ is the periodic function of $w$ given by (\ref{periodicchi}),
and has a f\/inite range given by (\ref{chimrange}). In this case the parametric solution given
by (\ref{u}) and (\ref{x}) is just a closed curve in the $(\chi,u)$ plane.
(A similar scenario was discussed in \cite[Section 3.3]{VP04} for the Degasperis--Procesi equation.)
This curve is symmetrical with respect to $\chi$ and has inf\/inite slope
at the two points where $u=v$.
Periodic composite weak solutions may be constructed from the closed curve.
(The notion of composite waves is discussed in \cite{S06,L05}, for example.)
For example, a periodic bell solution
with wavelength $4\chi_m$ is given in parametric form as follows:
\begin{gather}
u=u(w),\qquad
\chi=\begin{cases}
\chi(w){+}4j\chi_m,&\text{${}-w_m{+}2jK(m)\le w < w_m{+}2jK(m)$,}\\
\noalign{\vspace{4pt}}
\chi(w){+}(2+4j)\chi_m,&\text{$w_m{+}2jK(m)\le w < {-}w_m{+}(2{+}2j)K(m)$,}
\end{cases}\!\!\!\!
\label{composite}
\end{gather}
where $u(w)$, $\chi(w)$, $\chi_m$ and $w_m$ are given by (\ref{u}), (\ref{x}),
(\ref{chimrange}) and (\ref{wm}), respectively,
and $j=0,\pm1,\pm2,\ldots$.

\subsection[Waves with $m=1$]{Waves with $\boldsymbol{m=1}$}\label{sec4.3}

For $m=1$, the travelling-wave solution of the exROE is given by (\ref{usol}) and (\ref{xsol}).
From (\ref{psi1})--(\ref{psi2}), $\psi_1(1)=2$ and $\psi_2(1)=-1=\psi_3(1)$.
For all values of $\psi$, except for $\psi=-1$,
the range of $\chi$ as a function of $w$ is $(-\infty,\infty)$;
the corresponding possible solitary-wave prof\/iles for $(p+q)u$ may be categorized as follows:
\begin{center}
\begin{tabular}
{ll}
$\psi>2$:& smooth hump\\
$\psi=2$:& cuspon\\
$-1<\psi<2$:& loop\\
$\psi<-1$:& smooth hump\\
\end{tabular}
\end{center}
Furthermore, from (\ref{v}), $(p+q)v\gtrless 0$ according as $\psi\gtrless -c$.

When $\psi=-1$,
\begin{gather}
\chi=\frac{12k\tanh w}{p+q} ;
\label{xcorner}
\end{gather}
this can be obtained directly from (\ref{xsol}) or from (\ref{periodicchi}) with $m=1$.
Hence $\chi$ has a f\/inite range $[-\chi_m,\chi_m]$, where
\begin{gather}
\chi_m=\frac{12k}{|p+q|} ;
\label{xcornerm}
\end{gather}
this can be obtained directly from(\ref{xcornerm}) or from (\ref{chimrange}) with $m=1$.
By eliminating $\tanh w$ between (\ref{usol}) and (\ref{xcorner}), we obtain
\begin{gather}
u=u(\chi):=\frac{4k^2(c+2)}{p+q}-\frac{(p+q)\chi^2}{12} ,\qquad -\chi_m\le\chi\le\chi_m,
\label{compact}
\end{gather}
i.e.~a solitary wave with compact support.
Thus we can construct a composite weak solution for $(p+q)u$
in the form of spatially parabolic waves,
i.e.~corner waves, with discontinuous slope at the troughs, where
\begin{gather}
u=u(\chi-2j\chi_m),\qquad -\chi_m\le\chi-2j\chi_m\le\chi_m,\qquad j=0,\pm 1, \pm 2,\ldots.
\label{cornerwaves}
\end{gather}

The corner waves given by (\ref{cornerwaves})
are the maximum-amplitude
limit of a one-parameter family of periodic smooth-hump waves.
An example of such a family may be identif\/ied as follows.
When $\psi=-1$ and $0<m<1$, (\ref{v}) and (\ref{altpsi}) give the two possibilities
\begin{gather}
\left.
\begin{array}{ll}
c=+1:& \quad v=0,\quad \beta<0\quad\hbox{with $p$ and $q$ arbitrary};\\
c=-1:& \quad (p+q)v<0,\quad qv-\beta<0\quad\hbox{and}\quad (p-q)v+2\beta=0.
\end{array}
\right\}
\label{condition}
\end{gather}
Consider the one-parameter family of waves with parameter $m$ and for which
$p$, $q$, $\beta$ and $v$ are f\/ixed and satisfy either of the conditions in (\ref{condition}).
From Fig.~1 it can be seen that, for $0<m<1$, $\psi=-1$ lies below the lower curve
$\psi=\psi_3(m)$.
Hence, for $0<m<1$, the family are periodic smooth-hump waves.
From (\ref{kappa})--(\ref{u}),
we deduce that the amplitude of these waves is
\begin{gather}
\frac{3m|qv-\beta|}{|p+q|\sqrt{m^2-m+1}} .
\end{gather}
In the maximum-amplitude limit, i.e.~$m=1$, these waves become the corner
waves given by~(\ref{cornerwaves}) and have amplitude
\begin{gather}
\frac{3|qv-\beta|}{|p+q|} .
\end{gather}

\section{Examples}\label{sec5}

We illustrate the results in Section~\ref{sec4} by considering two examples.

\subsection{Example 1}\label{sec5.1}

Here we consider the simplest case, namely
the VE (for which $p=q=1$ and $\beta=0$).

In this case, $\psi$ given by (\ref{psi})
reduces to $\psi=c$, and $v\gtrless 0$ according as $\psi\gtrless -c$.
Hence, with $c=1$, we have $\psi=1$ and $v>0$. From Fig.~\ref{fig1} we deduce that the solution
comprises
periodic upright loops for $0<m<1$ and a solitary upright loop for $m=1$.
On the other hand, for $c=-1$, we have $\psi=-1$ and $v<0$.
From Fig.~\ref{fig1} we deduce that the solution comprises
periodic smooth humps for $0<m<1$ and a periodic corner-wave for $m=1$.
These are the results f\/irst given in~\cite{Va92,Pa93}.

\subsection{Example 2}\label{sec5.2}

Here we consider the GVE (for which $p=q=1$) with $c=1$ and arbitrary $\beta$.
(Note that the particular case for which $\beta=0$ is just the
VE with $c=1$ as discussed in Example 1.)

In this case, $\psi$ given by (\ref{psi}) reduces to{\samepage
\begin{gather}
\psi=1+\frac{\beta}{2\kappa^2},
\label{GVEpsi}
\end{gather}
and $v\gtrless 0$ according as $\psi\gtrless -1$.}

From (\ref{k}) and (\ref{GVEpsi}) with (\ref{psi1}), (\ref{psi2}) and (\ref{psi3}), the curves
$\psi=\psi_1(m)$, $\psi=\psi_2(m)$ and $\psi=\psi_3(m)$ correspond to
\begin{gather}
\beta= \beta_1(m):=
2k^2\left[m+1-\sqrt{m^2-m+1} \right],\label{beta1}\\
\beta= \beta_2(m):=
2k^2\left[m-2+\dfrac{3E(m)}{K(m)}-\sqrt{m^2-m+1} \right],\label{beta2}\\
\beta= \beta_3(m):=
2k^2\left[1-2m-\sqrt{m^2-m+1} \right]\label{beta3},
\end{gather}
respectively. Note also that $\psi=-1$ corresponds to
\begin{gather}
\beta=\beta_4(m):=-4k^2\sqrt{m^2-m+1}.
\label{beta4}
\end{gather}
so that
$v\gtrless 0$ according as $\beta\gtrless \beta_4(m)$.

\begin{figure}[t]
\centerline{\includegraphics[scale=0.87]{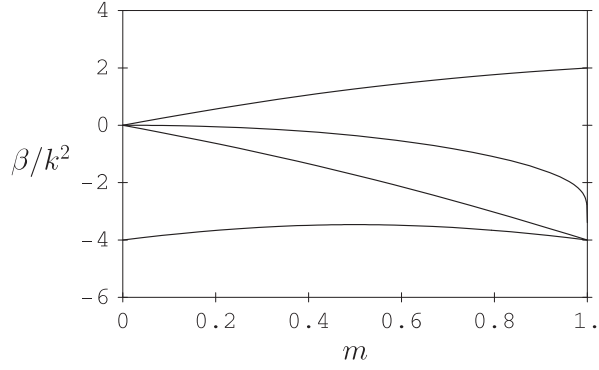}}
\vspace{-3mm}

\caption{From the top, the curves correspond to $\beta=\beta_1(m)$,
$\beta=\beta_2(m)$, $\beta=\beta_3(m)$ and $\beta=\beta_4(m)$, respectively.}\label{fig2}
\vspace{-1mm}
\end{figure}

The curves $\beta=\beta_1(m)$, $\beta=\beta_2(m)$, $\beta=\beta_3(m)$ and $\beta=\beta_4(m)$ are
plotted in Fig.~\ref{fig2}.

To illustrate the results in Section~\ref{sec4}, we let $k=1$ and consider the cases $m=0.5$ and $m=1$ separately.

\begin{figure}[t]
\centerline{\includegraphics{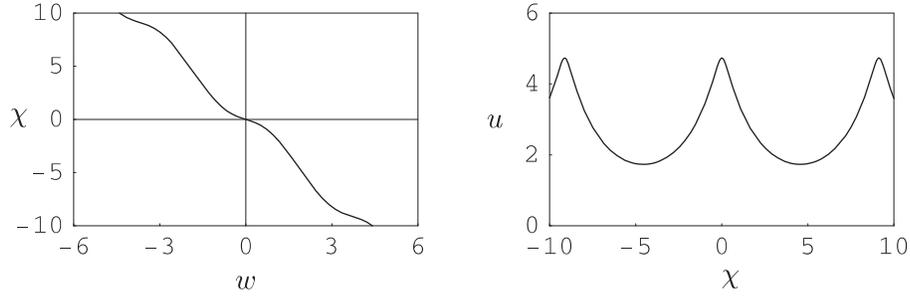}}
\vspace{-3mm}

\caption{$\beta=2.1>\beta_1$ so that $\psi>\psi_1$.
The wave prof\/ile comprises periodic smooth humps.}\label{fig3}
\vspace{-1mm}
\end{figure}

\begin{figure}[t]
\centerline{\includegraphics{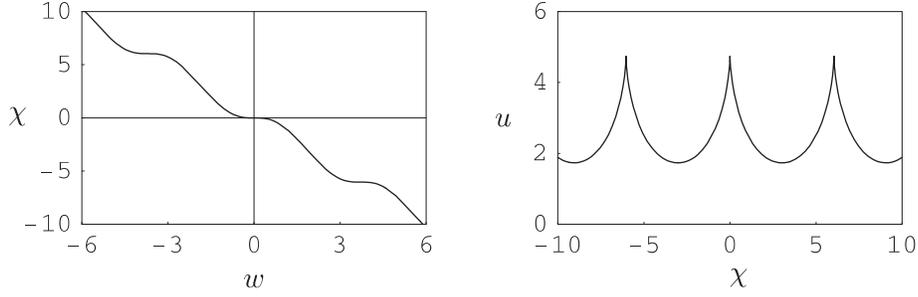}}
\vspace{-3mm}

\caption{$\beta=\beta_1$ so that $\psi=\psi_1$. The wave prof\/ile comprises periodic cuspons.}\label{fig4}
\vspace{-1mm}
\end{figure}

\begin{figure}[th!]
\centerline{\includegraphics{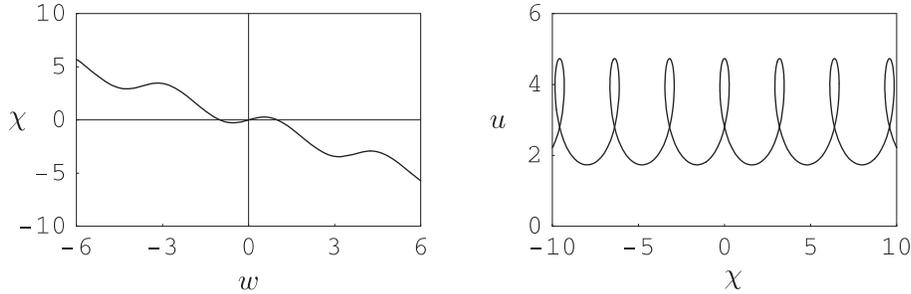}}
\vspace{-3mm}

\caption{$\beta=0.5$ so that $\beta_2<\beta<\beta_1$, i.e. $\psi_2<\psi<\psi_1$.
The wave prof\/ile comprises periodic loops.}\label{fig5}
\vspace{-1mm}
\end{figure}

\begin{figure}[th!]
\centerline{\includegraphics{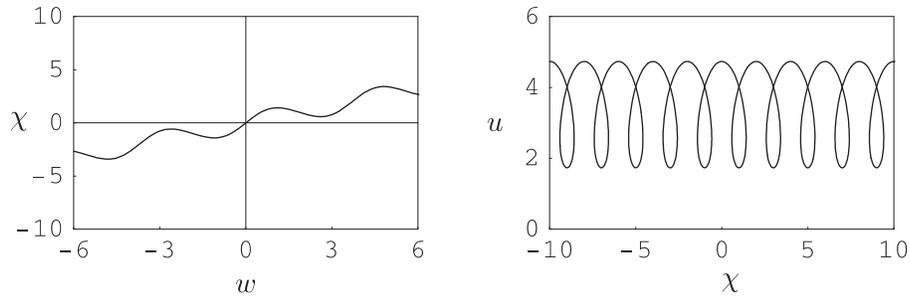}}
\vspace{-3mm}

\caption{$\beta=-0.9$ so that $\beta_3<\beta<\beta_2$, i.e.\ $\psi_3<\psi<\psi_2$.
The wave prof\/ile comprises periodic inverted loops.}\label{fig6}
\vspace{-1mm}
\end{figure}

\begin{figure}[th!]
\centerline{\includegraphics{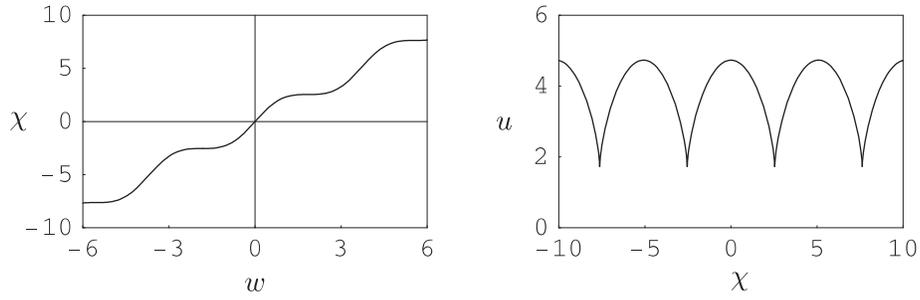}}
\vspace{-3mm}

\caption{$\beta=\beta_3$ so that $\psi=\psi_3$. The wave prof\/ile comprises periodic inverted cuspons.}\label{fig7}
\vspace{-1mm}
\end{figure}

\begin{figure}[th!]
\centerline{\includegraphics{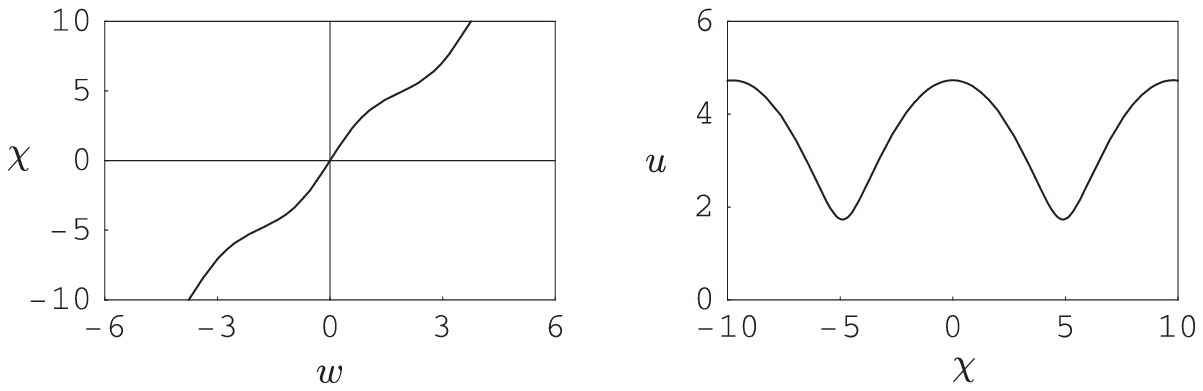}}
\vspace{-3mm}

\caption{$\beta=-3<\beta_3$ so that $\psi<\psi_3$.
The wave prof\/ile comprises periodic smooth humps.}\label{fig8}
\vspace{-1mm}
\end{figure}

\begin{figure}[th!]
\centerline{\includegraphics{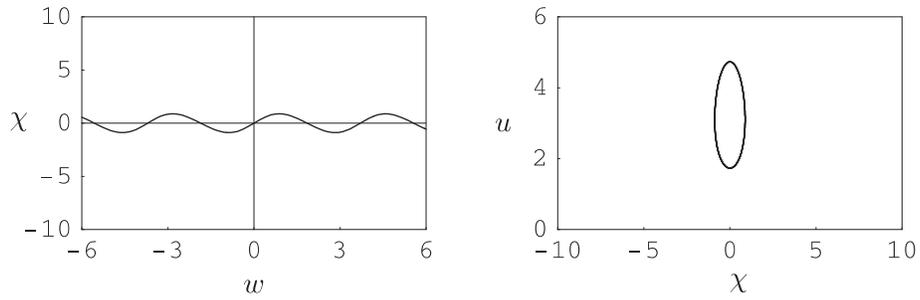}}
\vspace{-3mm}

\caption{$\beta=\beta_2$ so that $\psi=\psi_2$. The solution for $u$ is a closed curve.}\label{fig9}
\vspace{-1mm}
\end{figure}

\begin{figure}[th!]
\centerline{\includegraphics{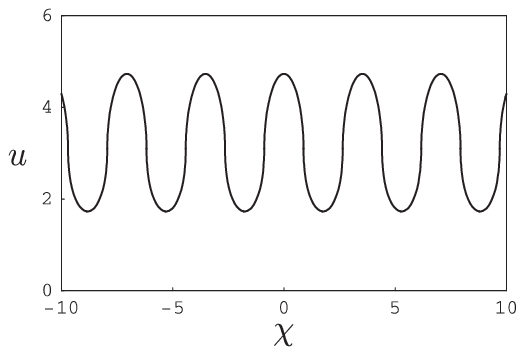}}
\vspace{-3mm}

\caption{A composite solution corresponding to Fig.~\ref{fig9}.
The wave prof\/ile comprises periodic bells.}\label{fig10}
\vspace{-1mm}
\end{figure}

With $k=1$ and $m=0.5$, (\ref{beta1})--(\ref{beta4}) give
$\beta_1\simeq 1.27$, $\beta_2\simeq -0.36$,
$\beta_3\simeq -1.73$ and $\beta_4\simeq -3.46$, respectively.
Figs.~\ref{fig3}--\ref{fig9} correspond to seven choices of $\beta$. In each f\/igure,
$\chi$ is plotted as an explicit function of $w$,
and $u$ is plotted as an implicit function of $\chi$.
In Figs.~\ref{fig3}--\ref{fig8}, the wave prof\/ile is periodic.
In Fig.~\ref{fig9}, the solution is a closed curve in the $(\chi,u)$ plane.
Fig.~\ref{fig10} illustrates the corresponding composite solution given by (\ref{composite}).

\begin{figure}[th!]
\centerline{\includegraphics{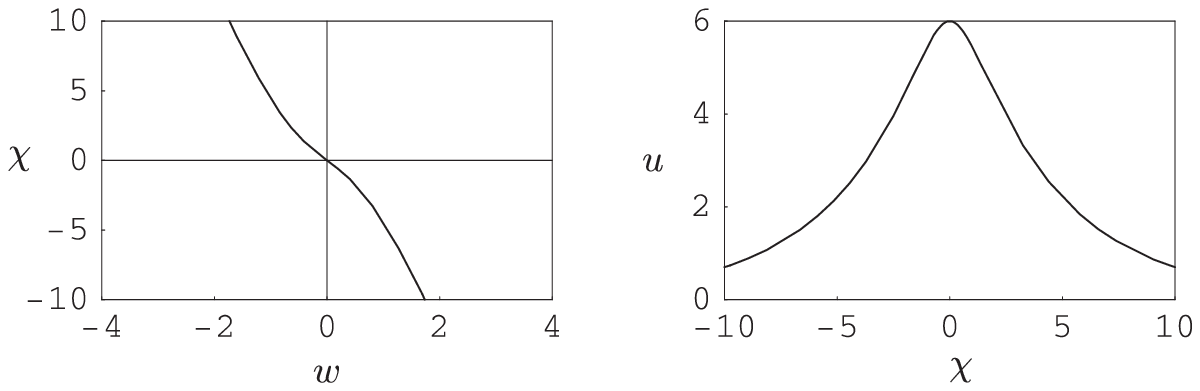}}
\vspace{-3mm}

\caption{$\beta=5$ so that $\psi>2$ and $v>0$.
The wave prof\/ile is a solitary smooth hump.}\label{fig11}
\vspace{-1mm}
\end{figure}

\begin{figure}[th!]
\centerline{\includegraphics{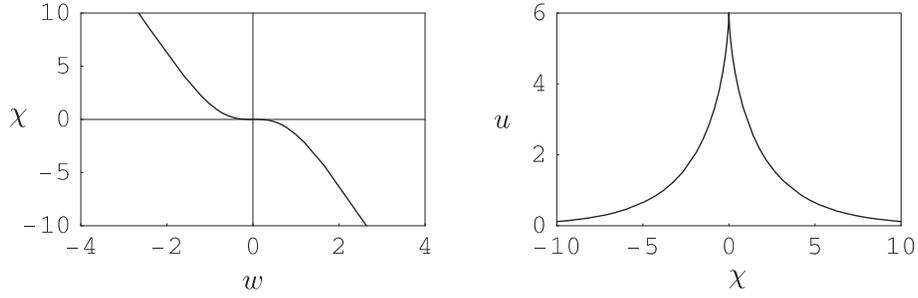}}
\vspace{-3mm}

\caption{$\beta=2$ so that $\psi=2$ and $v>0$. The wave prof\/ile is a solitary cuspon.}\label{fig12}
\vspace{-1mm}
\end{figure}

\begin{figure}[th!]
\centerline{\includegraphics{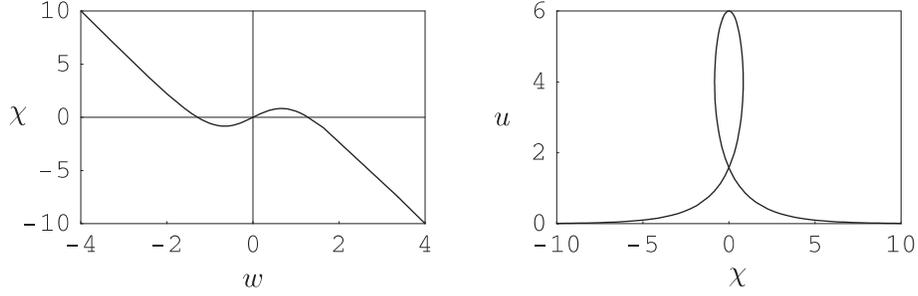}}
\vspace{-3mm}

\caption{$\beta=0$ so that $-1<\psi<2$ and $v>0$.
The wave prof\/ile comprises a solitary loop.}\label{fig13}
\vspace{-1mm}
\end{figure}

\begin{figure}[th!]
\centerline{\includegraphics{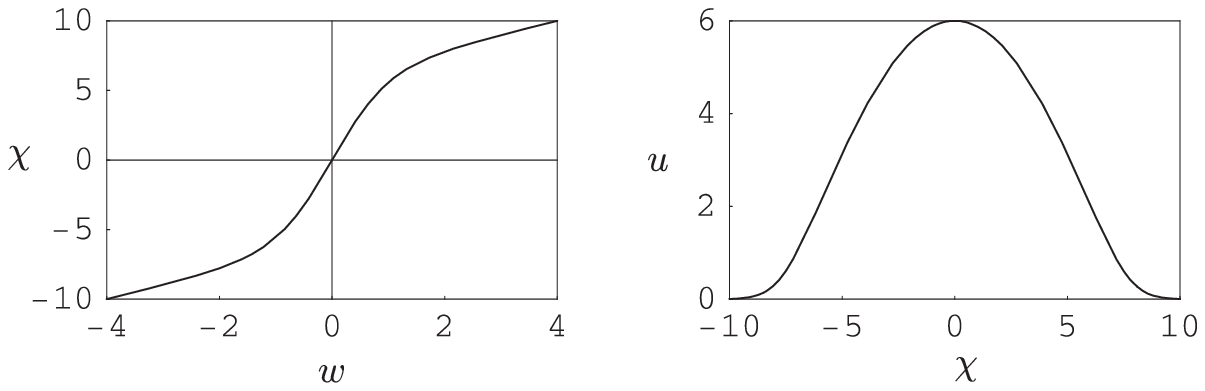}}
\vspace{-3mm}

\caption{$\beta=-5$ so that $\psi<-1$ and $v<0$.
The wave prof\/ile is a solitary smooth hump.}\label{fig14}
\vspace{-1mm}
\end{figure}

\begin{figure}[th!]
\centerline{\includegraphics{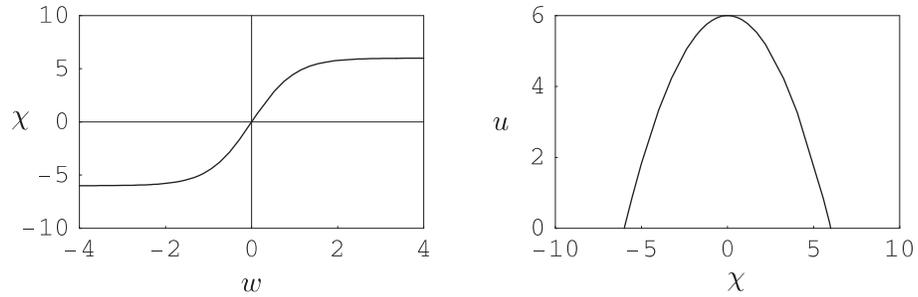}}
\vspace{-3mm}

\caption{$\beta=-4$ so that $\psi=-1$ and $v=0$. The solution for $u$ is a solitary wave with compact support.}\label{fig15}
\vspace{-3mm}
\end{figure}

\begin{figure}[th!]
\centerline{\includegraphics{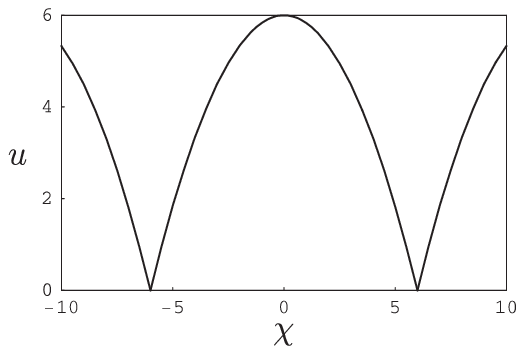}}
\vspace{-3mm}

\caption{A composite solution corresponding to Fig.~\ref{fig15}.
The wave prof\/ile comprises periodic corner waves.}\label{fig16}
\vspace{-3mm}
\end{figure}

With $k=1$ and $m=1$, (\ref{beta1})--(\ref{beta4}) give
$\beta_1=2$ and $\beta_2=\beta_3=\beta_4=-4$ as may be seen in Fig.~\ref{fig2}.
Figs.~\ref{fig11}--\ref{fig15} correspond to f\/ive choices of $\beta$.
In Figs.~\ref{fig11}--\ref{fig14}, the wave prof\/ile is a~solitary wave.
In Fig.~\ref{fig15}, the solution is the solitary wave with compact support given by (\ref{compact}).
Fig.~\ref{fig16} illustrates the corresponding composite solution comprising periodic corner waves
given by (\ref{cornerwaves}).
The solitary waves for $\beta/k^2\ne -4$ with $c=1$, i.e.~$v\ne 0$,
coincide with the single soliton solutions
given in \cite{MoPa01} for the GVE as derived via Hirota's method and
illustrated in Fig.~\ref{fig4} in~\cite{MoPa01}.
In \cite{MoPa01} it was shown that Hirota's method fails when $v=0$.
Now we know that, in this case, the solution is a solitary wave with
compact support or a periodic corner wave.

\section{Conclusion}\label{sec6}

We have found periodic and solitary travelling-wave solutions of an extended
reduced Ostrovsky equation.
These solutions were derived under the restrictions in (\ref{restrictions}).
We imposed these restrictions so that our solutions
reduce to solutions of the reduced Ostrovsky equation when $p=q=\pm 1$ and $\beta=0$.
We categorized the solutions by considering the values of~$\psi(m)$ and~$c$.

The def\/initions and derivations in Sections \ref{sec2}, \ref{sec3} and \ref{sec4} clearly require
the restrictions in (\ref{restrictions}) to hold. Of course, it is
of interest to ask what happens when these restrictions do not hold.
This question may be pursued elsewhere.
Here we record that additional bounded solutions do exist,
but no elegant categorization procedure appears to be possible.

Recently Li \cite{Li07} discussed the interpretation of multi-valued solutions of
several nonlinear wave equations. In Appendix B, we discuss brief\/ly how Li's interpretation
may be applied to the multi-valued waves derived in this paper. It turns out that
single-valued composite solutions may be constructed from our results.

Finally, we mention that we have applied Qiao's method \cite{ZQ07,Q08,QZ06}
to (\ref{poteqn}) in the form
\begin{gather}
(zz_{\eta})^2=f(z)
\end{gather}
in order to try to f\/ind an explicit solitary-wave solution for which
$z$ tends to a constant as $|\eta|\to\infty$. The method appears to give a negative result. This is
in agreement with our result in Section~\ref{sec4.3} where we showed that the only explicit
solitary-wave solution that arises from our method is the solitary wave with compact support
given by (\ref{compact}). The only explicit periodic-wave solution is the corner wave given by
(\ref{cornerwaves}).

\appendix
\pdfbookmark[1]{Appendix A}{appendixA}
\section*{Appendix A}
\setcounter{equation}{0}
\renewcommand{\theequation}{A.\arabic{equation}}

In this appendix we point out some errors in \cite{LLW07}.

Liu et al \cite{LLW07} claim that the exROE reduces to the VE when $p=\beta=0$ and $q=1$.
This is not true because the operators
$\cal D$ and $\frac{\partial}{\partial x}$ do not commute;
in fact they satisfy
\begin{gather}
\frac{\partial}{\partial x}\,{\cal D}\,\bullet =
\left(\frac{\partial u}{\partial x}+{\cal D}\right)\frac{\partial}{\partial x}\,\bullet.
\end{gather}
If these operators did commute, then the exROE with $p=\beta=0$ and $q\ne 0$ would reduce to
\begin{gather}
{\cal D}\left(\frac{\partial}{\partial x}{\cal D}u+qu\right)=0;
\label{dVE}
\end{gather}
this equation with $q=1$ is satisf\/ied by solutions of the VE and
so the claim in \cite{LLW07} would be correct.
The correct argument is as given in Section~\ref{sec1},
namely that when $p=q$ and $\beta=0$ the exROE reduces to equation~(\ref{diffROE}),
and equation~(\ref{diffROE}) with $p=1$ is
satisf\/ied by solutions of the~VE.

In \cite{LLW07}, the $qk^3$ term in equation~(3.18) should be $qk^2$ and consequently
$k^3$ should be $k^2$ in the
second term of the f\/irst and third equations in (3.21) and in the f\/ifth term of
the second equation in (3.21).
Incidentally, in \cite{LLW07} in order to get the f\/irst equation in (3.21),
$C$ in (3.18) has to be set to zero.
This is equivalent to setting $B=0$ in our (\ref{VEodeB}).

As a consequence of these corrections, the expressions for $\beta$ in (3.22) in
\cite{LLW07} should be
\begin{gather}
\beta=\frac{q\mp 4k^2c\sqrt{m^4-m^2+1}}{c}.
\label{beta}
\end{gather}
In (\ref{beta}), $c$ is not the $c$ in the present paper; it is equivalent to our $1/v$.
Liu et al \cite{LLW07} have adopted the convention used in \cite{BF71}
for the parameter of an elliptic function whereas
we have adopted the convention used in \cite{AS72};
it follows that $m^2$ in (\ref{beta}) is equivalent to $m$ in the present paper.
By noting this dif\/ference in notation and combining our (\ref{k}) with (\ref{beta}),
we see that~(\ref{beta}) is equivalent to our (\ref{kappa}).

Finally, $\xi_1$ in (3.25) in \cite{LLW07} should be
\begin{gather}
\xi_1=k\left[t-\left(\frac{q}{\beta+4k^2\sqrt{m^4-m^2+1}}\right)T\right]
\end{gather}
and $\xi_2$ in (3.27) in \cite{LLW07} should be
\begin{gather}
\xi_2=k\left[t-\left(\frac{q}{\beta-4k^2\sqrt{m^4-m^2+1}}\right)T\right].
\end{gather}
It follows that the $kq$ in (3.26) and (3.28) in \cite{LLW07} should be $q$.

The illustrative f\/igures for periodic waves in \cite{LLW07} are clearly incorrect;
they bear no resemblance
to the periodic humps, cusps or loops mentioned in the corresponding captions.

\pdfbookmark[1]{Appendix B}{appendixB}
\section*{Appendix B}
\setcounter{equation}{0}
\renewcommand{\theequation}{B.\arabic{equation}}

\begin{figure}[t]
\centerline{\includegraphics{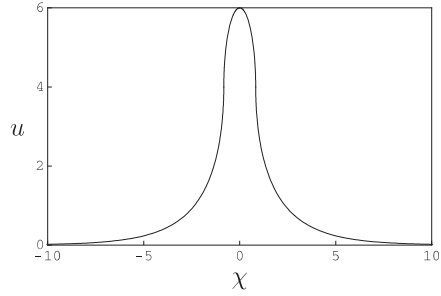}}
\vspace{-3mm}

\caption{A single-valued composite solution corresponding to Fig.~\ref{fig13}.}\label{fig17}
\vspace{-3mm}
\end{figure}

The wave prof\/iles for $u$ in Figs.~\ref{fig5}, \ref{fig6} and \ref{fig13} are multi-valued. Recently, Li \cite{Li07}
discussed the interpretation of similar solutions for other wave equations.
Here we will apply Li's ideas to the wave illustrated in Fig.~\ref{fig13}. The solution is given by
\begin{gather}
u=4(z+1),\qquad \chi=2\eta,
\end{gather}
where $z$ and $\eta$ are determined from
\begin{gather}
z^2_{\zeta}=-\frac{1}{3}\,(z+1)^2(2z-1)\qquad\hbox{and}\qquad\frac{d\eta}{d\zeta}=z
\end{gather}
so that
\begin{gather}
z=(1-3\tanh^2w)/2,
\label{z13}
\end{gather}
and
\begin{gather}
\eta=-2w+3\tanh w,\qquad w=\zeta/2,\qquad w\in (-\infty,\infty).
\end{gather}
Note that $\eta$ is not a monotonic function of $w$.
The phase portrait in the $(z,z_{\zeta})$-plane is a single
closed trajectory with a saddle point at $(-1,0)$.
However, the phase portrait in the $(z,z_{\eta})$-plane
consists of the stable and unstable manifolds through
the saddle in the region $-1\le z<0$, and an open curve
through $(1/2,0)$ for which $0<z\le 1/2$.
Li's point of view is that each of these three trajectories
corresponds to a dif\/ferent single-valued travelling-wave solution.
The multi-valued solution illustrated in Fig.~\ref{fig13} may be regarded as a
composite solution of these three single-valued solutions
but the three single-valued solutions may also be combined in dif\/ferent ways so as
to give a variety of composite single-valued solutions.
For example, with $z$ given by~(\ref{z13}) and $\eta$ given by
\begin{gather}
\eta=\begin{cases}
2w-3\tanh w - 2\eta_0,&w\in (-\infty,-w_0)\quad\hbox{so}\quad\eta\in (-\infty,-\eta_0)\\
-2w+3\tanh w,&w\in [-w_0,+w_0]\quad\hbox{so}\quad\eta\in [-\eta_0,+\eta_0]\\
2w-3\tanh w +2\eta_0,&w\in (w_0,\infty)\quad\hbox{so}\quad\eta\in (\eta_0,\infty),
\end{cases}
\end{gather}
where $w_0=\tanh^{-1}(1/3)$ and $\eta_0=-2w_0+\sqrt{3}$,
we have the wave illustrated in Fig.~\ref{fig17}. Note that, in this solution, $\eta$ is a
monotonic increasing function of the parameter $w$.
The corresponding periodic single-valued composite solution may be
constructed from the single-valued solutions making up the multi-valued waves in Figs.~\ref{fig5} and~\ref{fig6}.

\subsection*{Acknowledgements}
The author thanks the referees for some perceptive comments and for recommending
some additional references.

\pdfbookmark[1]{References}{ref}
\LastPageEnding

\end{document}